\documentclass[12pt,preprint]{aastex}
\usepackage{pdflscape}
\shorttitle{Search for exoplanets around northern circumpolar stars}
\shortauthors{Lee et al.}

\begin{document}

\title{Search for Exoplanets around Northern Circumpolar Stars\\
II. The Detection of Radial Velocity Variations in M Giant Stars HD~36384, HD~52030, and HD~208742}
\author{Byeong-Cheol Lee\altaffilmark{1,2}, Gwanghui Jeong\altaffilmark{1,2}, Myeong-Gu Park\altaffilmark{3}, Inwoo Han\altaffilmark{1,2},  David E. Mkrtichian\altaffilmark{4,5}, Artie P. Hatzes\altaffilmark{6}, Shenghong Gu\altaffilmark{7,8}, Jinming Bai\altaffilmark{7,8}, Sang-Min Lee\altaffilmark{1}, Hyeong-Il Oh\altaffilmark{3}, and Kang-Min Kim\altaffilmark{1}}

\altaffiltext{1}{Korea Astronomy and Space Science Institute 776, Daedeokdae-ro, Yuseong-gu, Daejeon 305-348, Korea; bclee@kasi.re.kr}
\altaffiltext{2}{Astronomy and Space Science Major, Korea University of Science and Technology, Gajeong-ro Yuseong-gu, Daejeon 305-333, Korea}
\altaffiltext{3}{Department of Astronomy and Atmospheric Sciences, Kyungpook National University, Daegu 702-701, Korea}
\altaffiltext{4}{National Astronomical Research Institute of Thailand, Chiang Mai 50200, Thailand}
\altaffiltext{5}{Crimean Astrophysical Observatory, Taras Shevchenko National University of Kyiv, Nauchny, Crimea, 98409, Ukraine}
\altaffiltext{6}{Th{\"u}ringer Landessternwarte Tautenburg (TLS), Sternwarte 5, 07778 Tautenburg, Germany}
\altaffiltext{7}{Yunnan Observatories, Chinese Academy of Sciences, Kunming 650011, China}
\altaffiltext{8}{Key Laboratory for the Structure and Evolution of Celestial Objects, Chinese Academy of Sciences, Kunming 650011, China}

%---------------------------------------------------------------------------------
\begin{abstract}
We present the detection of long-period RV variations in HD~36384, HD~52030, and HD~208742 by using the high-resolution, fiber-fed Bohyunsan Observatory Echelle Spectrograph (BOES) for the precise radial velocity (RV) survey of about 200 northern circumpolar stars. Analyses of RV data, chromospheric activity indicators, and bisector variations spanning about five years suggest that the RV variations are compatible with planet or brown dwarf companions in Keplerian motion. However, HD~36384 shows photometric variations with a period very close to that of RV variations as well as amplitude variations in the weighted wavelet Z-transform (WWZ) analysis, which argues that the RV variations in HD~36384 are from the stellar pulsations. Assuming that the companion hypothesis is correct, HD~52030 hosts a companion with minimum mass 13.3 $\it M_\mathrm{Jup}$ orbiting in 484 days at a distance of 1.2 AU. HD~208742 hosts a companion of 14.0 $\it M_\mathrm{Jup}$ at 1.5 AU with a period of 602 days. All stars are located at the asymptotic giant branch (AGB) stage on the H-R diagram after undergone the helium flash and left the giant clump.With stellar radii of 53.0\,$R_{\odot}$ and 57.2\,$R_{\odot}$ for HD~52030 and HD~208742, respectively, these stars may be the largest yet, in terms of stellar radius, found to host sub-stellar companions. However, given possible RV amplitude variations and the fact that these are highly evolved stars the planet hypothesis is not yet certain.
\end{abstract}


\section{INTRODUCTION}
The difference between a planet and a brown dwarf is still under debate. One distinction can be made according to different formation mechanisms.
Brown dwarf companions are thought to be formed by gravitational collapse in protostellar clouds like stellar binary systems (Bate 2000)
or the gravitational instability in protostellar disks (Boss 2000; Rice et al. 2003), similar to the formation mechanism for massive planet.
However, it is difficult to determine this boundary from observations of specific objects.
Another criterion to classify a planet versus a brown dwarf is based on nuclear fusion inside the object.
Burrows et al. (1997) used the criterion that a planet occupies the mass range below the deuterium burning limit of $\sim$13\,$\it M_\mathrm{Jup}$, between this and the mass for the onset of core hydrogen fusion ($\sim$80\,$\it M_\mathrm{Jup}$) objects are brown dwarfs. However, this mass cutoff is difficult to define precisely because the deuterium fusion depends to some extent on the composition of the object such as a helium abundance, an initial deuterium abundance, and a metallicity (Spiegel et al. 2011).

Recently, it was suggested that the upper limit to planet masses should be set at  20$-$30\,$\it M_\mathrm{Jup}$ (Lovis \& Mayor 2007; Sahlmann et al. 2011; Schneider et al. 2011). On the other hand, Hatzes \& Rauer (2015) proposed that all objects in the mass range 0.3$-$60\,$\it M_\mathrm{Jup}$ are giant planets based on the fact that in the mass-density diagram there are no distinguishing features in density between giant planets and brown dwarfs. The Exoplanet Database\footnote{http://exoplanets.org/, http://exoplanetarchive.ipac.caltech.edu/} lists planets up to 24$-$25\,$\it M_\mathrm{Jup}$, but that is an arbitrary boundary (Schneider et al. 2011).

Generally, planet or brown dwarf companion surveys using the precise radial velocity (RV) technique are best suited for G and K type stars. Moreover, the surveys were conducted mostly around main-sequence (MS) stars, and there is a lack of statistical data for giant stars.
%It is well known that brown dwarfs are very common in the field (Reid et al. 1999), in open clusters (Bouvier et al. 1998; Mart{\'{\i}}n et al. 1998), in close companions to other brown dwarfs and to very low mass stars (Bouy et al. 2003; Close et al. 2003). Nevertheless,  initial the long-term RV survey of Campbell et al. (1988) and Murdoch et al. (1993) did not reveal any brown dwarf companions to nearby solar-type stars despite sufficient instrumental sensitivity.
RV surveys have established that  there is a distinct lack of brown dwarfs within 3 -- 4 AU of the solar-type stars and this  generally referred to as a brown dwarf desert (Marcy \& Butler 2000). Despite planet search surveys having monitored thousands of stars so far, few such objects have been discovered around evolved stars: 14 $\it M_\mathrm{Jup}$ for HD 13189 (Hatzes et al. 2005); 19.8 $\it M_\mathrm{Jup}$ for NGC 4349 No. 127 (Lovis \& Mayor 2007); 19.4 $\it M_\mathrm{Jup}$ for 11 Com (Liu et al. 2008), and 37.6 $\it M_\mathrm{Jup}$ for HD 119445 (Omiya et al. 2009).
%Also, Borgniet et al. (2017) present an occurrence rate of brown dwarfs and planetary companions and to compare the results with evolved stars and lower-mass Main-Sequence stars.
In recent studies, exoplanets have been detected with a frequency of $\sim$5\% for MS stars and $\sim$10\% for giant stars, whereas brown dwarfs are estimated to be around less than 1\% of Sun-like stars within $\sim$3 AU (Kraus et al. 2008) and around 2\% for 1 -- 3 $M_{\odot}$ within  the period of $\sim$1000 days (Borgniet et al. 2017).
While exoplanets are relatively common as stellar companions, brown dwarfs seem to be rare. Finding more brown dwarfs in the desert can provide a key information to star and planet formation because it represents the extreme mass limit of both processes.
%Furthermore, brown dwarfs are rarely found near evolved stars. This dramatic dichotomy indicates that perhaps the formation of brown dwarfs and stellar companions involves very different mechanisms.
%The formation of brown dwarfs is under dispute  and it is unclear whether brown dwarfs could form via planetary formation processes. Low-mass stars and brown dwarfs can form through the fragmentation of dense filaments and disks, possibly followed by early ejection from these dense environments, which truncates their growth in mass (Bonnell et al. 2007).

To study the formation mechanisms of planets or brown dwarfs using the RV method, observations of stars with various spectral and luminosity types are imperative.
Although several precise RV surveys of evolved stars have been conducted (Frink et al. 2002; Hatzes et al. 2005; Setiawan et al. 2005; Johnson et al. 2007; Jones et al.2011; Lovis \& Mayor 2007; Niedzielski et al. 2007; Sato et al. 2007; Han et al. 2010; Wittenmyer et al. 2011), the existence of substellar companions in giant stars is still not clearly understood.
Finding more planets can provide an important clue to star and planet formation because they represent the extreme mass limit of both processes.

In this paper, we present three RV variations around giant stars from the survey. In Sect. 2, we describe the observational data and reduction procedures. We give the stellar characteristics in Sect. 3, and the RV analysis and the nature of the RV measurements for each star are put in Sect. 4. Finally, in Sect. 5, we summarize the result from this study.

%________________________________________________________________

\section{OBSERVATIONS AND DATA REDUCTION}
We conducted the “\textbf{S}earch for \textbf{E}xoplanets around \textbf{N}orthern circumpolar \textbf{S}tars (SENS; see Lee et al. 2015)”, which uses the 1.8~m telescope at Bohyunsan Optical Astronomy Observatory (BOAO) in Korea.
About 200 circumpolar stars have been selected for the survey, of which 5\% are dwarfs, 40\% giants, and 55\% stars of unclassified luminosity class covering all spectral types.

Precise RV observations were carried out using the fiber-fed high-resolution Bohyunsan Observatory Echelle Spectrograph (BOES; Kim et al. 2007) attached to the 1.8 m telescope at BOAO. Among three kinds of available resolutions, R = 90 000, 45 000, and 27 000, the optical fiber of 45 000 is most suitable due to trade off between efficiency and resolution. We used an iodine absorption (I$_{2}$) cell to obtain precise RV measurements (Valenti et al. 1995; Butler et al. 1996), which superimposes thousands of molecular absorption lines over the object spectra in the wavelength region of 4900$-$6100 ${\AA}$. An exposure time is limited to less than about 20 minutes to reduce the line broadening from differential barycentric motion and it provides an average signal to noise (S/N) of $\sim$150.

Basic data reduction was carried out using the IRAF software and the precise RV measurements related to the I$_{2}$ analysis were undertaken using the RVI2CELL (Han et al. 2007). The RV standard star $\tau$ Ceti has been monitored since 2003 to check the long-term stability of the BOES. Over this time the star has shown an rms scatter of $\sim$7 m~s$^{-1}$ (Lee et al. 2013).

%________________________________________________________________
%
\section{STELLAR CHARACTERISTICS}
\subsection{Fundamental parameters}
The basic stellar parameters were taken from the \emph{HIPPARCOS} catalog (ESA 1997). The luminosities came from the result of Anderson \& Francis (2012), that are computed based on V-band photometry with a bolometric correction supplied by Masana et al. (2006).
The stellar atmospheric parameters of temperature ($T_{\mathrm{eff}}$), metal abundance ($\mathrm{[Fe/H]}$), surface gravity (log $\it g$), and micro turbulence ($v_{\mathrm{micro}}$) were determined using the TGVIT (Takeda et al. 2005). This was based on the Kurucz (1992) atmosphere models using around 130$-$150 equivalent width (EW) measurements of Fe~I and Fe~II lines for each star, which usually require three constraint conditions of excitation equilibrium, ionization equilibrium, and matching the curve of growth shape (Takeda et al. 2002).

HD 52030 has a spectral type of K0 III (ESA 1997). The \emph{B$-$V} color index and the $T_{\mathrm{eff}}$, however, indicate that the star could  be a late K or early M-type giant. The \emph{V$-$K} color of 4.08 is more consistent with an M1~III (\emph{V$-$K} = 4.05) rather than a K0 III (\emph{V$-$K} = 2.31) according to Bessel \& Brett (1988).
HD~208742 listed as spectral type of K5 with unclassified luminosity class (ESA 1997). The star could be classified as a giant according to the value of the low surface gravity and a large stellar radius compared to that of the dwarf star (see Section 3.2). The \emph{HIPPARCOS} color index of approximately 1.6 and our estimation of the $T_{\mathrm{eff}}$ indicate that the spectral type of HD~208742 is close to a late K or early M. The \emph{V$-$K} color of 3.94 is similar to an M0.5~III in Bessel \& Brett (1988). As a result, HD~208742 should be an early M giant star.
The basic stellar parameters are summarized in Table~\ref{tab1}.

\subsection{Stellar radius and mass}
Stellar mass and stellar radius are the major factors in determining the fate of companions in the evolution track of giant stars. We determined the stellar radii and masses from the stellar  positions in the color--magnitude diagram with using the theoretical stellar isochrones of Bressan et al. (2012). Bayesian estimation method (J{\o}rgensen \& Lindegren 2005; da Silva et al. 2006) was also adopted.
We derived $R_{\star}$ = 38.4 $\pm$ 3.4\,$R_{\odot}$ and $M_{\star}$ = 1.14 $\pm$ 0.15\,$M_{\odot}$ for HD~36384, $R_{\star}$ = 53.0 $\pm$ 6.1\,$R_{\odot}$ and $M_{\star}$ = 1.09 $\pm$ 0.16\,$M_{\odot}$ for HD~52030, and $R_{\star}$ = 57.2 $\pm$ 7.1\,$R_{\odot}$ and $M_{\star}$ = 1.18 $\pm$ 0.22\,$M_{\odot}$ for HD~208742.

%__________________________________________________________________

\subsection{Rotational period}
As stars evolve away from the MS, they become cooler and rotate slower.
Low-amplitude and long-period RV variations in evolved stars may arise from the rotational modulation of surface features such as star spots (Lee et al. 2008, 2012).
In order to estimate the stellar rotational velocities, we used a line-broadening model (Takeda et al. 2008) and the automatic spectrum-fitting technique (Takeda 1995). We estimated the minimum rotational velocities of 4.5 km~s$^{-1}$ for HD~36384, 4.1 km~s$^{-1}$ for HD~52030, and 5.4 km~s$^{-1}$ for HD~208742. Based on the rotational velocities and the stellar radii, we derived the upper limits for the rotational period of 440 $\pm$ 40 days for HD~36384, 650 $\pm$ 80 days for HD~52030, and 540 $\pm$ 70 days for HD~208742.

%__________________________________________________________________
%
\section{RADIAL VELOCITY VARIATIONS AND THEIR ORIGINS}
\subsection{Orbital solutions}
\begin{figure}
   \centering
   \includegraphics[width=8cm]{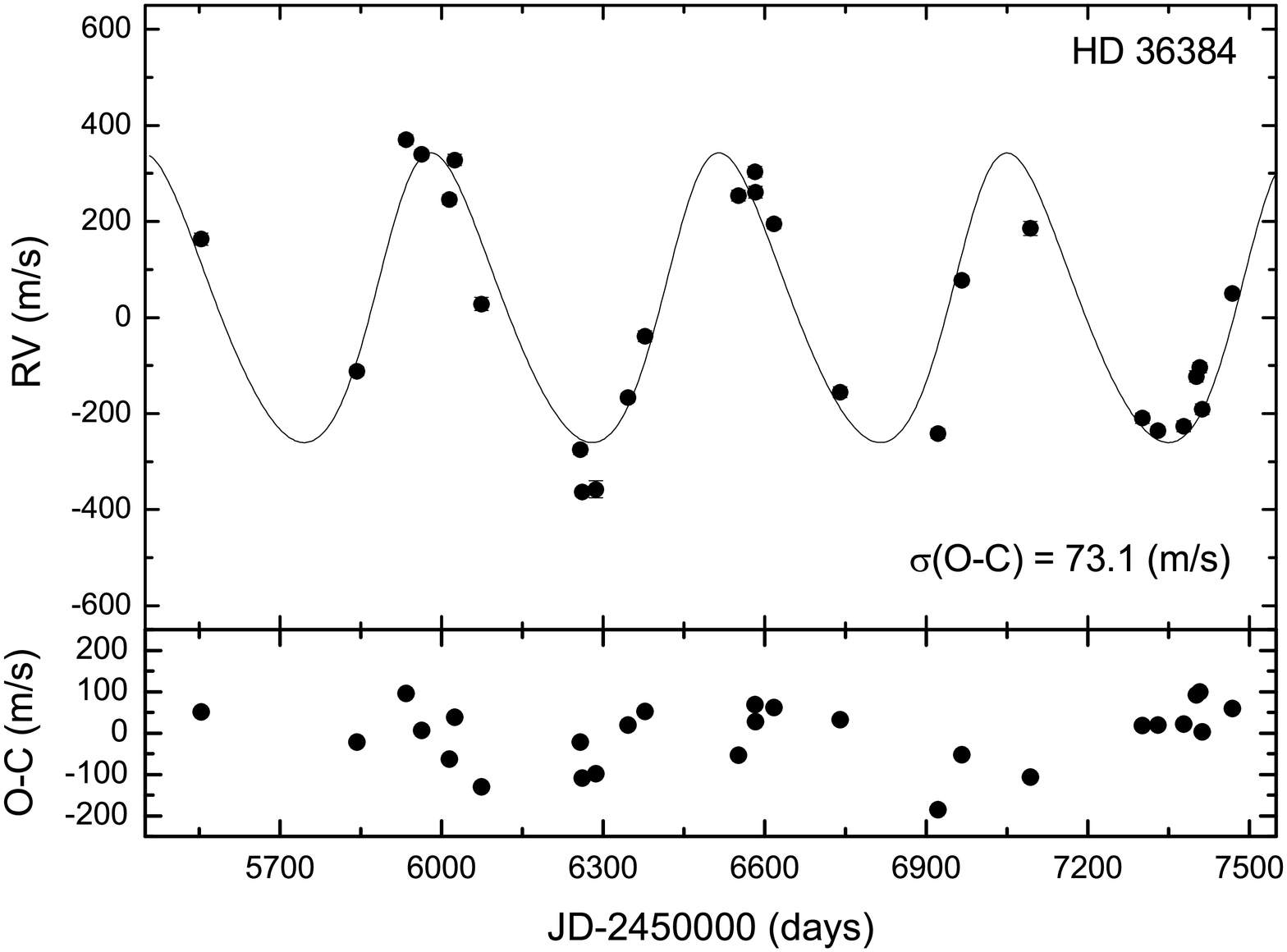}
      \caption{RV measurements for HD 36384 from December 2010 to March 2016. (\emph{Top panel}) Observed RVs for HD 36384, and the solid line is a Keplerian orbital fit with a period of 535.1 days, a semi-amplitude of 302.1 m s$^{-1}$, and an eccentricity of 0.13, yielding a minimum companion mass of 13.1 $M_{\rm{Jup}}$. (\emph{Bottom panel}) Residual velocities remained after subtracting the Keplerian orbital fit from observations.
              }
         \label{rv1}
   \end{figure}

\begin{figure}
   \centering
   \includegraphics[width=8cm]{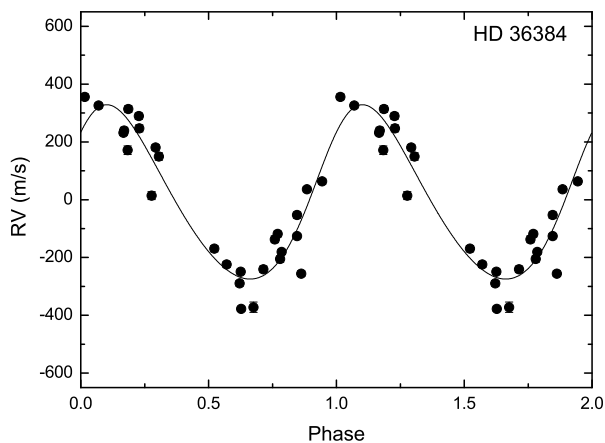}
      \caption{RV variations for HD 36384 phased with the orbital period of 535.1 days.
        }

        \label{phase1}
   \end{figure}

We obtained 27 BOES spectra for HD~36384 spanning five years, which show clear periodic variations (Fig.~\ref{rv1}).  The  Lomb-Scargle (L-S) periodogram (Scargle 1982) results in a peak of 535.1 days. Figure~\ref{phase1} shows the RV data phased to the orbital period superposed on the Keplerian fit. The L-S power of this peak corresponds to a false alarm probability (FAP) of $<$ 1 $\times$ 10$^{-5}$  (top panel in Fig.~\ref{power1}), adopting the procedure described in Cumming (2004)\,
which is based on the goodness-of-the-fit statistic for the difference between the Keplerian fit and the linear fit to the data.
The FAP is determined by a Monte Carlo method in which fake data sets containing noise only are compared with the observed power or from the analytic distribution of power (Cumming 2004).
We found a semi-amplitude $K$ = 302.1 $\pm$ 20.4 m s$^{-1}$ and an eccentricity $e$~= 0.13 $\pm$ 0.12 for a Keplerian orbital fit.

\begin{figure}
   \centering
   \includegraphics[width=8cm]{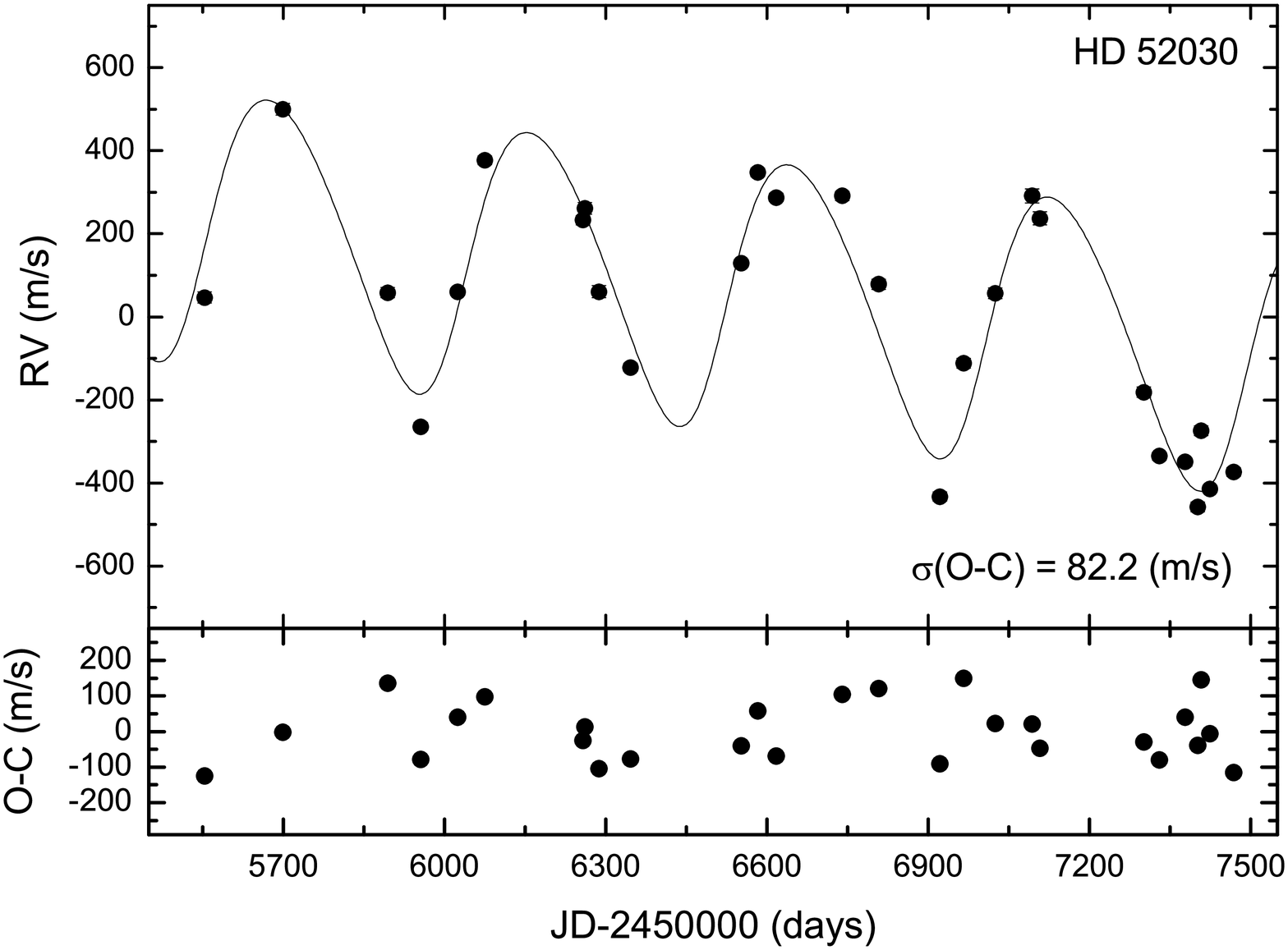}
      \caption{RV measurements for HD 52030 from December 2010 to March 2016. (\emph{Top panel}) Observed RVs for HD 52030, and the solid line is a linear plus Keplerian orbital fit with a period of 484.3 days, a semi-amplitude of 329.0 m s$^{-1}$, and an eccentricity of 0.14, yielding a minimum companion mass of 13.3 $M_{\rm{Jup}}$. (\emph{Bottom panel}) Residual velocities remained after subtracting the Keplerian orbital fit from observations.
              }
         \label{rv2}
   \end{figure}

\begin{figure}
   \centering
   \includegraphics[width=8cm]{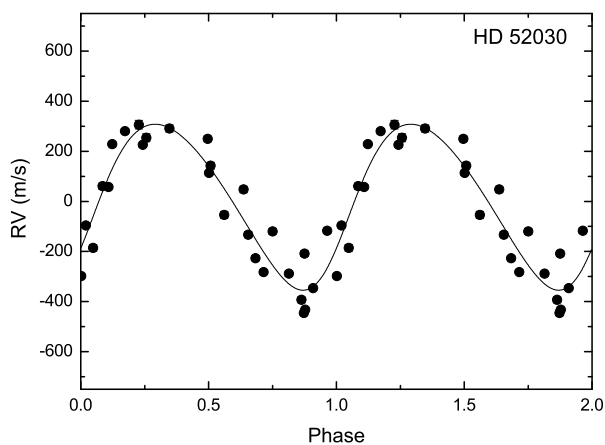}
      \caption{RV variations for HD 52030 phased with the orbital period of 484.3 days.
        }

        \label{phase2}
   \end{figure}

\begin{figure}
   \centering
   \includegraphics[width=8cm]{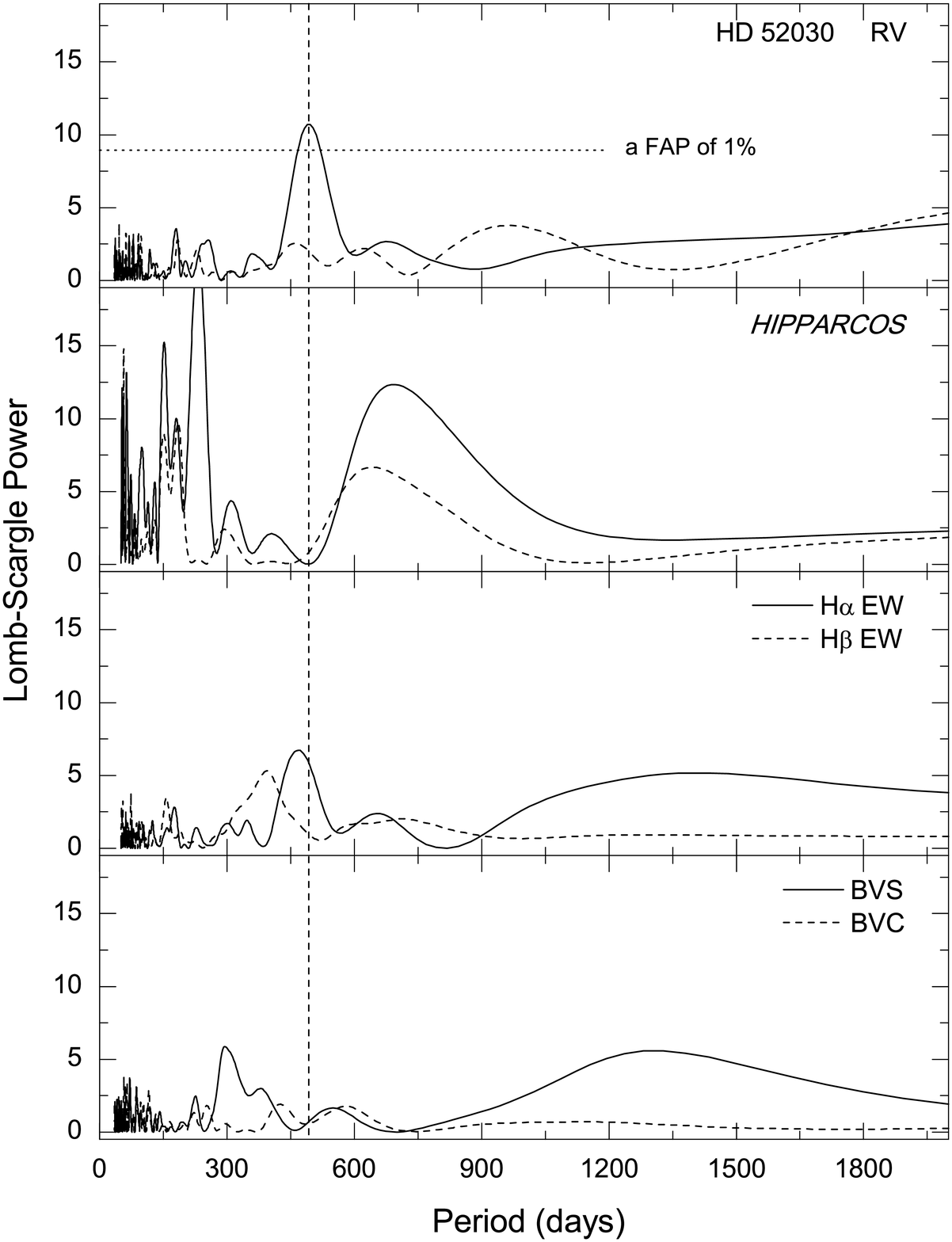}
      \caption{L-S periodograms of the RV measurements, the \emph{HIPPARCOS} photometric measurements, the EW variations of the hydrogen lines, and the bisector variations for HD 52030 (\emph{top} to \emph{bottom panel}). The vertical dashed line indicates the location of the period of 484 days.
      (\emph{Top panel}) The solid line is the L-S periodogram of the RV measurements for five years, and the periodogram shows a significant power at a period of 484.3 days. The dashed line is the periodogram of the residuals after removing the main period fit from the original data. The horizontal dotted line indicates a FAP threshold of 1 $\times 10^{-2}$ (1\%).
        }
        \label{power2}
   \end{figure}

Observations for HD~52030 with the BOES took place between December 2010 and March 2016. We gathered a total of 27 spectra during five and half years (Fig.~\ref{rv2} and Table~\ref{tab3}).
Figure~\ref{rv2} shows more obvious linear secular change, which may come from unknown long-term period binary.
The RV measurements phased to the orbital period are shown in Fig. ~\ref{phase2}. The L-S periodogram shows a peak at 484 days (Fig.~\ref{power2}). The RV variations could be fitted with a Keplerian orbit with a period of 484.3 $\pm$ 4.2 days, a semi-amplitude of 329.0 $\pm$ 21.8 m s$^{-1}$, and an eccentricity of 0.14 $\pm$ 0.14.

\begin{figure}
   \centering
   \includegraphics[width=8cm]{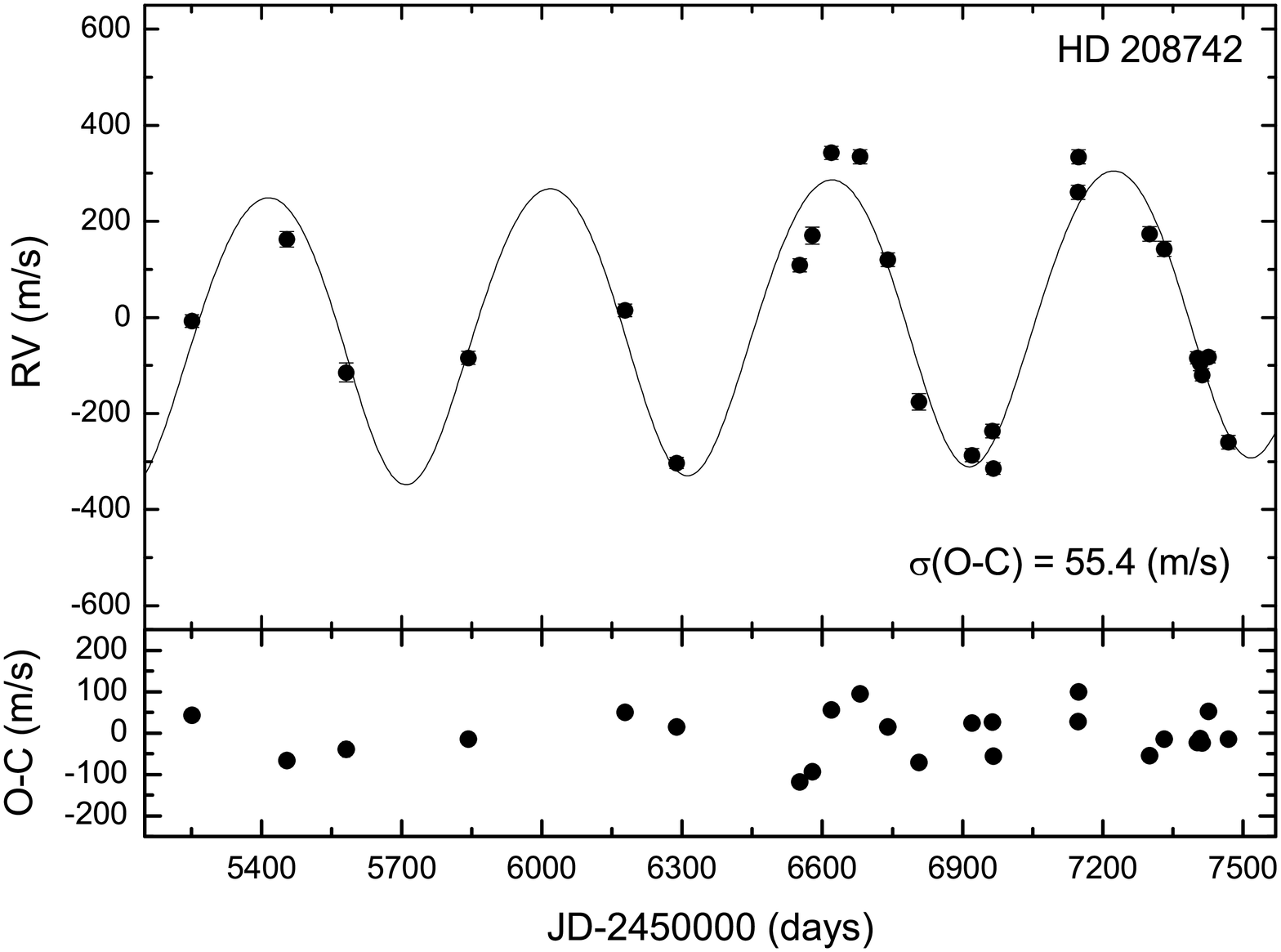}
      \caption{RV measurements for HD 208742 from February 2010 to March 2016. (\emph{Top panel}) Observed RVs for HD 208742, and the solid line is a Keplerian orbital fit with a period of 602.8 days, a semi-amplitude of 303.0 m s$^{-1}$, and an eccentricity of 0.06, yielding a minimum companion mass of 14.2 $M_{\rm{Jup}}$. (\emph{Bottom panel}) Residual velocities remained after subtracting the Keplerian orbital fit from observations.
              }
         \label{rv3}
   \end{figure}

\begin{figure}
   \centering
   \includegraphics[width=8cm]{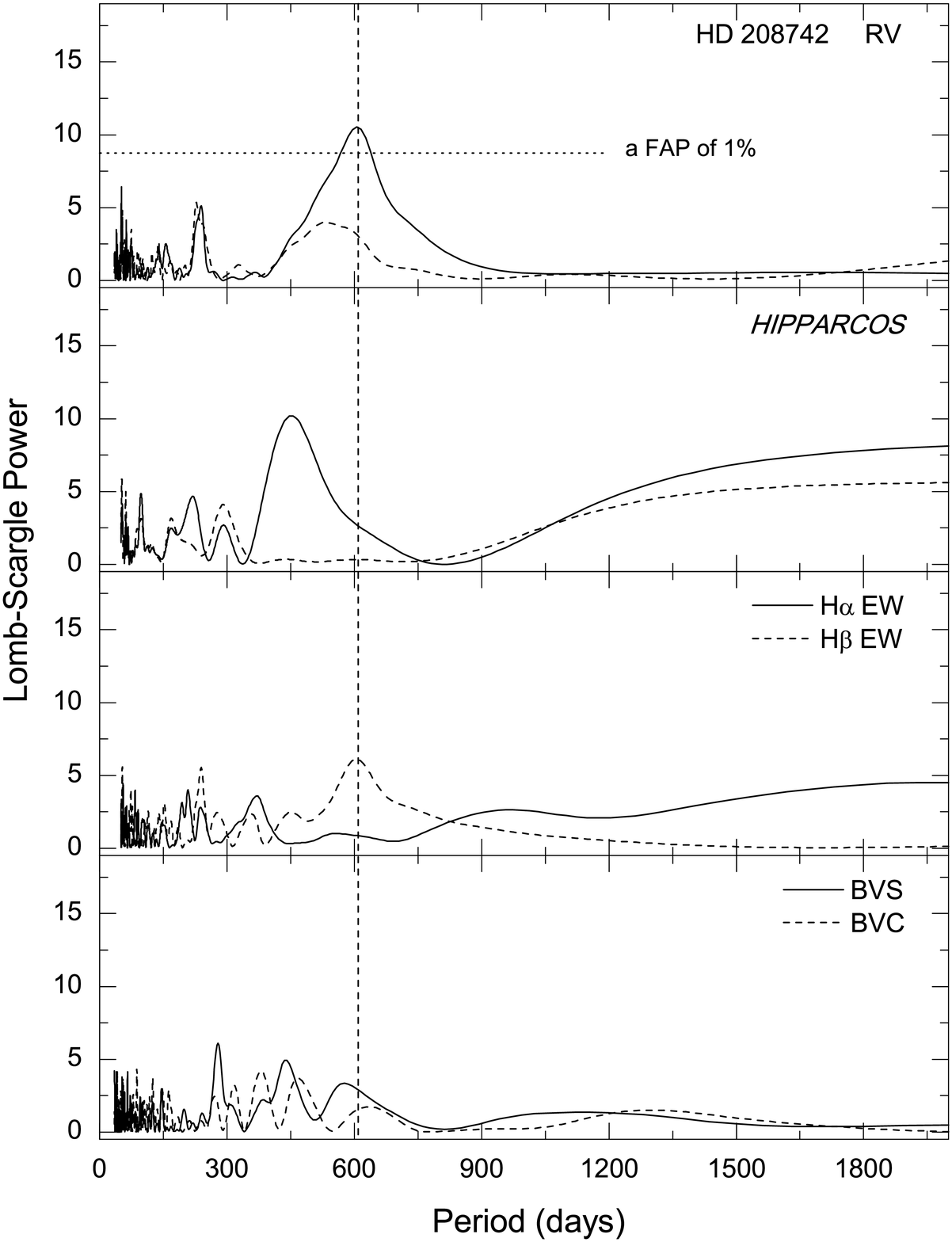}
      \caption{L-S periodograms of the RV measurements, the \emph{HIPPARCOS} photometric measurements, the EW variations of the hydrogen lines, and the bisector variations for HD 208742 (\emph{top} to \emph{bottom panel}). The vertical dashed line indicates the location of the period of 602 days.
      (\emph{Top panel}) The solid line is the L-S periodogram of the RV measurements for five years, and the periodogram shows a significant power at a period of 602.8 days. The dashed line is the periodogram of the residuals after removing the main period fit from the original data. The horizontal dotted line indicates a FAP threshold of 1 $\times 10^{-2}$ (1\%).
        }
        \label{power3}
   \end{figure}

Since February 2010, we have gathered 24 spectra for HD~208742 as displayed in Fig.~\ref{rv3} and listed in Table~\ref{tab4}.  The L-S periodogram of the RV measurements for HD~208742 shows a significant peak at a period of 602.8 days (top panel in Fig.~\ref{power3}). The L-S power of this peak corresponds to a FAP of less than 1~$\times$~10$^{-4}$. We found semi-amplitude $K$ = 303.0 $\pm$ 16.6 m~s$^{-1}$ and eccentricity $e$~= 0.06 $\pm$ 0.13 for a Keplerian fit.

The residual RV variations after removing the Keplerian orbital fit are 73.1~m~s$^{-1}$ for HD~36384, 82.1 m~s$^{-1}$ for HD~52030, and 55.4 m~s$^{-1}$ for  HD~208742. They are significantly larger than the RV precision for the RV standard star $\tau$~Ceti ($\sim$7~m~s$^{-1}$) and the typical internal error of individual RV precisions of $\sim$10.9~m~s$^{-1}$ (HD~36384), $\sim$11.7~m~s$^{-1}$ (HD~52030), and $\sim$14.4~m~s$^{-1}$ (HD~208742), respectively.
We searched for periodic variations in the residual RV measurements using the L-S periodogram.
These show no additional periodic signals (dashed lines in the top panels of Figs.~\ref{power1}, \ref{power2}, and \ref{power3}). Large residual RVs can be the typical features of evolved stars (Setiawan et al. 2003; Hatzes et al. 2005; D{\"o}linger et al. 2007; de Medeiros et al. 2009), which tend to increase toward later type stars (Hekker et al. 2006).
They are most likely to arise from stellar oscillations. These can be stochastic so they may not appear as periodic signals, particularly due to the sparse sampling of the data. The scaling relationships of Kjeldsen \& Bedding (1995) yield expected RV amplitudes from stellar oscillations of $\sim$80 m~s$^{-1}$ for HD~36384, $\sim$186 m~s$^{-1}$ for HD~52030, and $\sim$194 m~s$^{-1}$ for HD~208742 comparable or greater than the observed residuals after removing the Keplerian fit.
Thus, these residuals may be related to the stellar oscillations.

\subsection{Photometric variations}
To search for possible brightness variations that might be caused by the rotational modulation of large star spots, we analyzed the \emph{HIPPARCOS} photometry data. The data were obtained from December 1989 to February 1993 (HD~36384), from December 1989 to March 1993 (HD~52030), and from December 1989 to February 1993 (HD~208742): a total of 177, 226, and 141 \emph{HIPPARCOS} measurements were made for HD~36384, HD~52030, and HD~208742, respectively. All stars showed rms scatters of 0.012$-$0.014 mag. This corresponds to a variation over the time span of the observations of 0.19\%, 0.22\%, and 0.18\% for the three stars, respectively.
The L-S of the photometric variations for the stars are shown in Figs.~\ref{power1}, \ref{power2}, and \ref{power3}. All three stars exhibit certain level of periodic variations. Especially, HD~36384 show a periodic photometric variations close to the 535-d period seen in RV variations.

\subsection{Surface activity}
Large spots, produced by the chromospheric activity, crossing the stellar surface can create a spectroscopic signature that might be misinterpreted as a presence of companion. When searching for exoplanets using the RV technique, thus, the possibility of variations caused by chromospheric activity must be considered.
The EW variations of Ca II H \& K, H$_{\beta}$, Mg~I~b triplet, He~I~D$_{3}$, Na~I~D$_{1}$ \& D$_{2}$, H$_{\alpha}$, and Ca II infrared triplet (IRT) lines are in general used as optical chromospheric activity indicators.
Of these, the Na I D$_{1}$ \& D$_{2}$ and Mg~I~b triplet lines are excluded from the study due to their location inside I$_{2}$ molecular region. The Ca~II~IRT lines are not suitable either due to significant fringing and saturation of our CCD spectra at wavelengths longer than $\sim$7500~{\AA}.

\begin{figure}
   \centering
   \includegraphics[width=7cm]{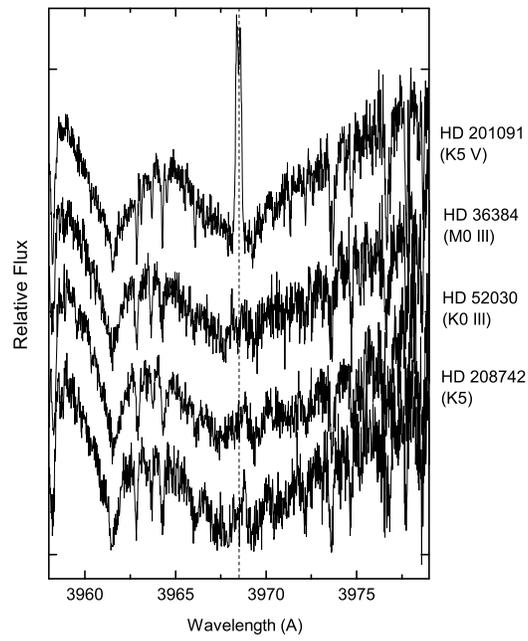}
      \caption{Ca II H line cores for our program stars including the chromospheric active star HD~201091. There are hints of weak core reversals in the center of the Ca II H line for three stars. The vertical dotted line indicates the center of the Ca II H regions.
        }
        \label{Ca1}
   \end{figure}

The most common indicators of chromospheric activity are the Ca II H \& K lines, which are well studied for F to K type stars (Montes et al. 1995; K{\"u}rster et al. 2003; Lee et al. 2015; Hatzes et al. 2015).
However, the BOES does not provide a sufficient S/N ratio to estimate EW variations in the Ca~II~H \& K line region.
We used the Ca~II~H line as this spectral region was of higher quality than for the Ca~II~K feature.
Figure~\ref{Ca1} shows the Ca II H line region of four BOES spectra, and the chromospherically active star HD~201091 (K5 V) is shown for comparison.
Unfortunately, the spectra are of too poor quality to resolve any emission feature that might be present.
%Unfortunately, all spectra are not clear enough to resolve the emission feature.
%There may be a slight central emission indicating a low level of stellar activity.
Possible weak features are visible in the center of Ca~II~H line for all three stars, although not strong enough to guarantee the presence of the chromospheric activity in the sample.
We thus measured variations in H$_{\alpha}$ and H$_{\beta}$ EW for the sample.
Specifically, the H$_{\alpha}$ line is sensitive to the chromospheric structure and conditions in giant stars (Cram \& Mullan 1985; Smith et al. 1989).
Recently, Lee et al. (2016) found that two hydrogen lines of $\mu$~UMa show periodic variations ($\sim$473 days) consistent with the secondary RV period of 471 days, whose origin is probably a stellar activity.
We measured the EWs using a band pass of $\pm$ 1.0 ${\AA}$ for H$_{\alpha}$ and $\pm$ 0.8 ${\AA}$ for H$_{\beta}$ centered on the core of the lines to avoid nearby blending lines and ATM H$_{2}$O absorption lines.
The individual chromospheric activity measurements of the H lines in L-S periodograms are shown in the third panel of Figs.~\ref{power1}, \ref{power2}, and \ref{power3}. They do not show any significant periodic variations.

\subsection{Bisector variations}
In order to investigate other causes that may produce apparent RV variations, such as stellar rotational modulations or pulsations rather than the orbital motion, a spectral line shape analysis was performed using high-resolution stellar lines (Hatzes, Cochran, \& Bakker 1998; Queloz et al. 2001). Two bisector quantities were calculated: the velocity span (BVS) and the velocity curvature (BVC). The former is the velocity difference between the two flux levels of the bisector and the latter is the difference of the velocity span of the upper half and lower half of the bisectors.

To avoid contaminations, unblended spectral features with high flux levels and free of the I$_{2}$ absorption lines were chosen.
The BOES iodine cell has negligible absorption lines beyond about 6200 {\AA}.
So, we selected 11 spectral lines of  V~I~6251.8, Fe~I~6262.6, Ti~I~6325.2, Cr~I~6330.1, Fe~I~6335.3, Fe~I~6411.6, Fe~I~6421.4, Ca I 6499.7, Fe I 6574.2, Fe I 6593.9, and Ni I 6767.8 {\AA}.
The difference between the bisectors of the profiles at flux levels of 0.8 and 0.4 of the central depth were used as velocity spans, which thereby avoid the spectral core and wing where errors of the bisector measurements are large.
The average BVS and BVC were computed using the bisector measurements of all 11 spectral lines after subtracting the mean value for each spectral line. Figure~\ref{power1}, \ref{power2}, and \ref{power3} (\emph{bottom panel}) show L-S periodograms of BVS and BVC for each star, respectively. None of them exhibit any meaningful periodic variations.

\section{RESULTS}
\begin{figure}
   \centering
   \includegraphics[width=8cm]{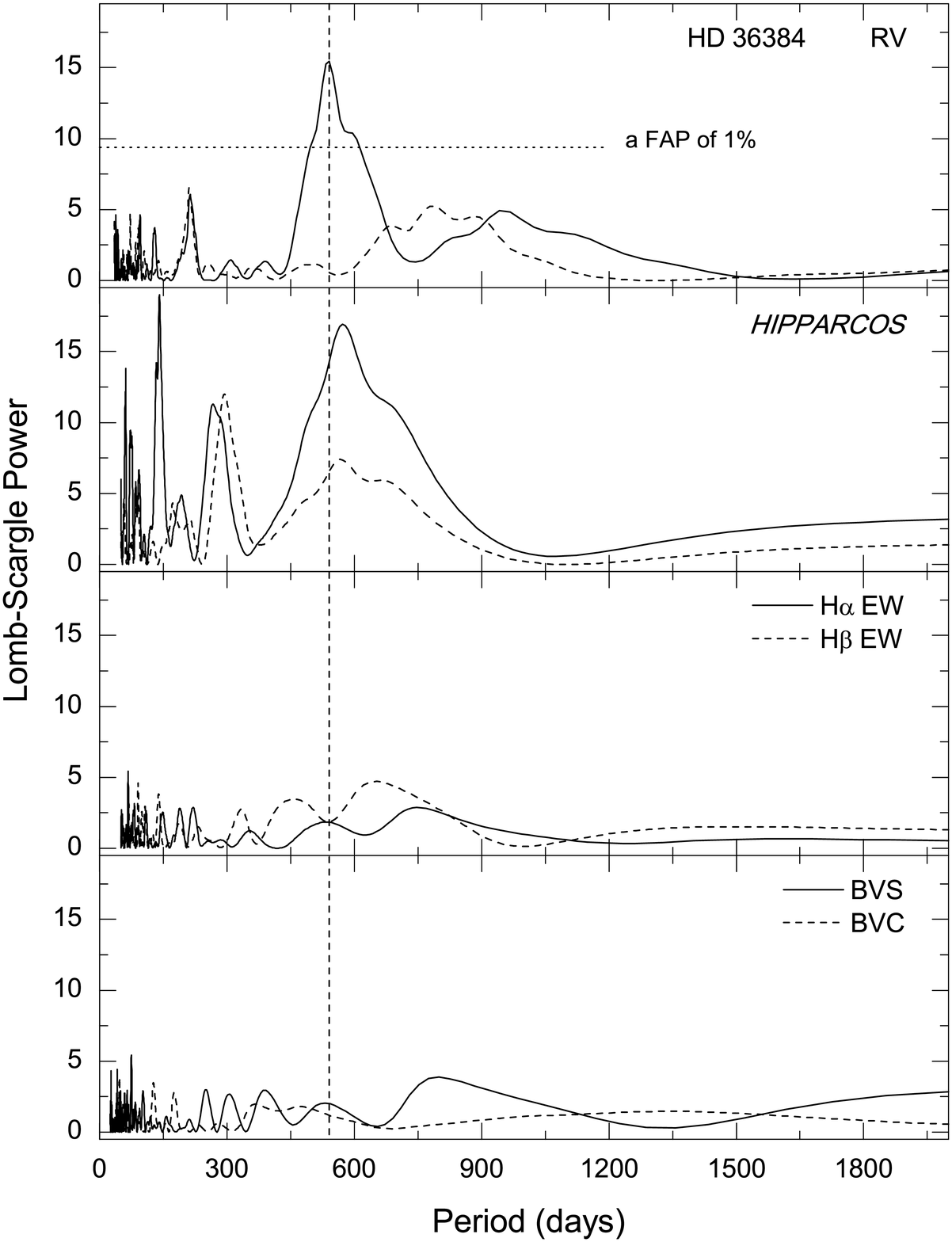}
      \caption{L-S periodograms of the RV measurements, the \emph{HIPPARCOS} photometric measurements, the EW variations of the hydrogen lines, and the bisector variations for HD 36384 (\emph{top} to \emph{bottom panel}). The vertical dashed line indicates the location of the period of 535 days.
      (\emph{Top panel}) The solid line is the L-S periodogram of the RV measurements for five years, and the periodogram shows a significant power at a period of 535.1 days. The dashed line is the periodogram of the residuals after removing the main period fit from the original data. The horizontal dotted line indicates a FAP threshold of 1 $\times 10^{-2}$ (1\%).
        }
        \label{power1}
   \end{figure}

We found periodic RV variations in HD~36384, HD~52030, and HD~208742.
For HD~36384, the EW variations of the hydrogen lines were examined for any RV fluctuations that would be induced by stellar chromospheric activity. Figure~\ref{power1} shows the L-S periodogram of the H$_{\alpha}$ and H$_{\beta}$ EW variations (third panel). No correlation was found between the hydrogen line EWs and the RV variations. This means that HD~36384 exhibited at most a modest chromospheric activity at the time of observations.
Two kinds of line bisector variations were measured and there is no correlation between the RV and the bisector variations (bottom panel in Fig.~\ref{power1}).
Finally, the \emph{HIPPARCOS} photometric data were analyzed for brightness variations in HD~36384 (second panel in Fig.~\ref{power1}). The L-S periodogram shows large power at $\sim$60, $\sim$140, $\sim$270, and $\sim$570 days. The last period is close to the RV period of 535.1 $\pm$ 5.3 days.
A planetary companion is bound to produce low-amplitude RV variations, the amplitude of which will not vary in time.
Therefore, the variation of the amplitude over time can be another diagnostic tool to disprove the companion hypothesis.

\begin{figure}
   \centering
   \includegraphics[width=8cm]{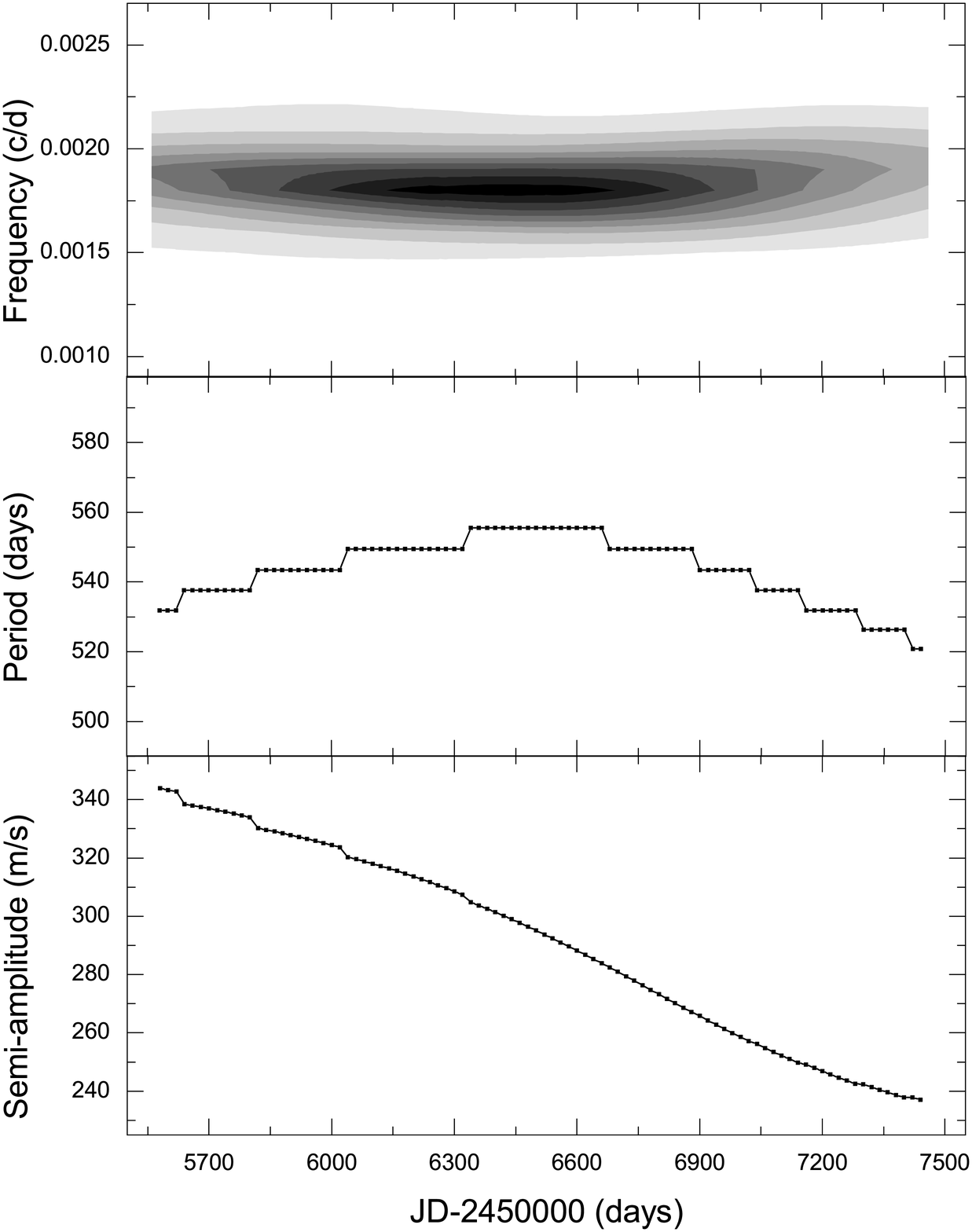}
      \caption{The wavelet analysis for HD~36384. The likelihood map of the frequency, the most likely period, and the semi-amplitude of the main frequency component are shown (\emph{top} to \emph{bottom panel}).}
        \label{wwz}
   \end{figure}

We have applied to our RV data the weighted wavelet Z-transformation (WWZ) developed by Foster (1996). Wavelet analysis is well-suited to detect transient periodic fluctuations, as well as changes in their parameters, because it can focus attention on a limited time span of the data. However, wavelet transform can often be susceptible to false fluctuations for uneven sampling. By utilizing the test statistic that is similar to the chi-square statistic, follows the usual F-distribution but is less sensitive to the effective number of data, WWZ can detect the time evolution of the period and amplitude even for unevenly sampled time series (Foster 1996). Figure~\ref{wwz} shows WWZ, and the maxima period has barely changed over the observations.
However, the amplitude of the RV variation decreased by more than 30\%. These results suggest that RV variations are not caused by a companion.

The L-S periodograms of the \emph{HIPPARCOS} measurements, the EW variations of the hydrogen lines, and line bisector variations for HD~52030 are shown in Fig.~\ref{power2}.
Although the \emph{HIPPARCOS} measurements show significant peaks at $\sim$60, $\sim$160, $\sim$230, and $\sim$680 days, they are irrelevant for the RV variations of 484 days. Furthermore, the lack of any significant peaks in the L-S periodograms of chromospheric activity indicators and bisector variations supports that the RV variations of HD~52030 are caused by a companion.
By adopting a stellar mass $M_{\star}$ = 1.09 $\pm$ 0.16 $M_{\odot}$ for HD~52030, we obtain a companion mass \emph{m}~sin~$i$ = 13.3 $\pm$ 2.3 $M_{\rm{Jup}}$ and a semi-major axis $\it{a}$ = 1.2 $\pm$ 0.1 AU (Table~\ref{tab5}).

In Figure~\ref{power3}, the dominant peak of the \emph{HIPPARCOS} measurements for HD~208742 is near 450 days, significantly different from the RV period of 602 days. A peak in the L-S periodogram of H$_{\beta}$ nearly at 600 days has the FAP of about 10\%, statistically not meaningful. The periodograms of the bisector also show no periodicity near the 602-day period of the RV variation. All these results together lead us to conclude that the RV variations observed in HD~208742 are not ascribed to an intrinsic stellar process but to an orbiting companion.
Adopting a mass $M_{\star}$ = 1.18 $\pm$ 0.22 $M_{\odot}$, the companion has a mass \emph{m}~sin~$i$ = 14.0 $\pm$ 2.0 $M_{\rm{Jup}}$ at a distance of 1.5 $\pm$ 0.2 AU from HD~208742 (Table~\ref{tab5}).

%__________________________________________________________________

\section{DISCUSSION}
The spectral types for two program stars are reclassified with the results from the literature and our new estimation of the stellar parameters. HD~52030, previously known as a K0 type, is changed to an M1 type and HD~208742, previously known as a K0 with unknown luminosity class, is reclassified to an early M giant.
Generally, most giants have intrinsic RV variations and exhibit pulsations and/or surface activities. To determine the nature of the RV variabilities requires comprehensive consideration of all relevant analyses.

\subsection{Pulsations}
Analyses of rotational periods, H line EW variations, and line bisector measurements of all three stars do not show any obvious evidence of the RV variations caused by a rotational modulation of surface features.
However, several strong periods were found in the \emph{HIPPARCOS} photometric data, which suggests the possibility of stellar pulsations. In particular, the period of $\sim$570 days in HD~36384 close to 535 day period in RV variations strengthens the possibility.
We further checked for any changes in the amplitude of RV variations as a function of time for HD~36384. Although the sampling of of the RV data is not sufficient for this analysis, the WWZ analysis hints a secular decrease of the amplitude of RV variations over five years. In addition, HD~36384 is classified as a variable star in the SIMBAD database. Even though the mechanism to produce such low-amplitude, long-period pulsation modes is not well understood, we suspect that the current RV variations seen in HD~36384 is more likely to be the result of stellar pulsations rather than an orbiting companion.

\subsection{Fake planets}
There is recent evidence that long-period variations in K giant stars that mimic a planet signal may in fact
be intrinsic stellar variability (Reffert et al. 2015; Hatzes, \emph{private communication}).
Such RV amplitude change is obviously not compatible with orbiting companions, and the periodic RV variations in such systems are more likely due to the long-period stellar pulsations, known to occur in giant stars.
These findings warn us that the usual diagnostic tests for orbiting planets such as comparison of variations of chromospheric activity indicators and BVS/BVC against those of RV spanning less than 3 to 5 years may not be conclusive as previously believed, especially for giant stars. Follow up observations over longer time scale will help to select out such `fake planet' systems.

\subsection{The fate of companions on the evolution track}
\begin{figure}
   \centering
   \includegraphics[width=8cm]{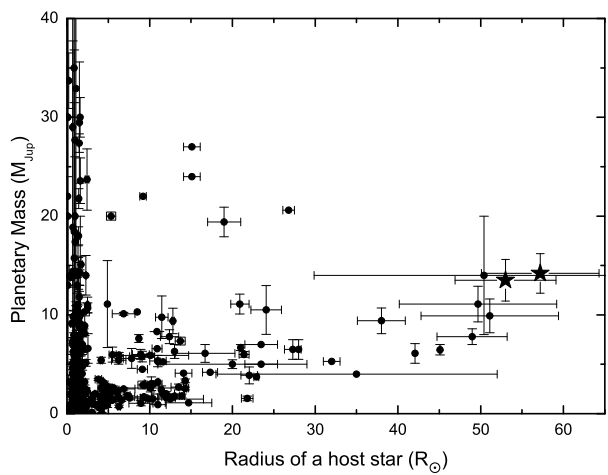}
      \caption{Distribution of candidate planetary companions in the stellar radius and in the planetary mass as of September 2016. Closed circles are known companions (http://exoplanets.org) and pentagrams on the far right two new candidate planetary companions HD~52030~b and HD~208742~b of this work \emph{(left to right)}.}
        \label{stat}
   \end{figure}

The physical parameters from the Keplerian fit for HD~52030~b, and HD~208742~b place the minimum mass of the companions in the planetary mass region. Figure~\ref{stat} shows the distribution of the stellar radii versus the planetary masses.
Of the $\sim$ 3000 host stars harboring companions with known stellar radii, 97\% are smaller than subgiant phase of 5\,$R_{\odot}$ ($\sim$0.025 AU; within an orbit of a 3-day hot Jupiter). The stars crossing the subgiant branch do not undergo significant expansion.
The largest host stars discovered so far to have companions are HD~208527 (51.1\,$R_{\odot}$) and HD~13189 (50.39\,$R_{\odot}$). HD~52030 and HD~208742 in our study may be the largest stars with a substellar companion at a radius of 53.0\,$R_{\odot}$ (0.25 AU) and 57.2\,$R_{\odot}$ (0.27 AU), respectively.
Approximately 65\% of companions resided inside 0.25 AU from host stars and most companions around dwarfs will undergo an ineluctable destiny during the evolutionary process.

The fate of planets around evolved stars may be dramatically altered as the host star evolves off the MS stage.
The companions may have been engulfed  or moved outward by the expanding host stars when the latter increases its radius as it ascends the red giant branch (RGB) and the asymptotic giant branch (AGB).
When the Sun climbs the AGB stage with its largest radial extent of 213\,$R_{\odot}$ (0.99 AU), however, the Earth would move out to 1.69 AU, narrowly escaping being engulfed only in the presence of mass loss (Sackmann et al. 1993).
Recent studies showed that planets with masses smaller than 1~$M_{\rm{Jup}}$ do not survive during the planetary nebula phase
if located initially at an orbital distance closer than 3$-$5 AU (Villaver \& Livio 2007; Kunitomo et al. 2011; Mustill \& Villaver 2012; Mustill et al. 2014), whereas planets more massive than 2~$M_{\rm{Jup}}$ survive down to orbital distances of $\sim$3 AU.
For 1.5$-$5 \,$M_{\odot}$, planets in orbits within 0.5 AU around RGB stars could have been engulfed by the host stars at the tip of the RGB due to the tidal force from the host stars (Sato et al. 2008).
Johnson et al. (2007) also considered the possibility that the lack of close-in planets around giants may be attributable to engulfment by the expanding atmospheres of the host stars.

\begin{figure}
   \centering
   \includegraphics[width=8cm]{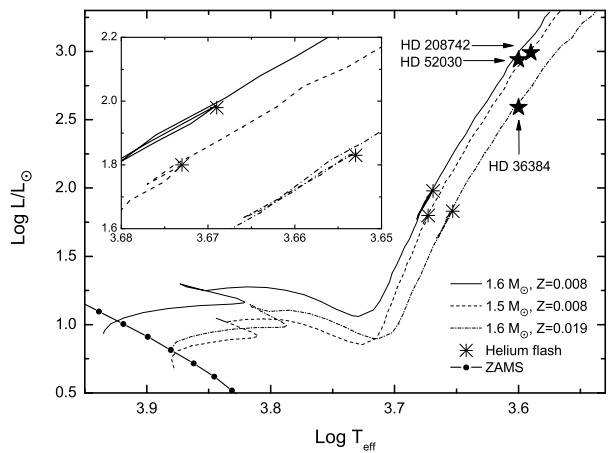}
      \caption{The H$-$R diagram illustrating the properties of the stars (pentagrams) presented in this paper. Pairs of evolutionary tracks from Girardi et al. (2000) for stars with 1.6\,$M_{\odot}$, Z = 0.008 (solid line), 1.5\,$M_{\odot}$, Z = 0.008 (dashed line), and 1.6\,$M_{\odot}$, Z = 0.019 (dash dotted line). Asterisks mean the helium flash point. Small box is an enlarged view of the helium flash region.}
        \label{hrd}
   \end{figure}

Figure~\ref{hrd} shows HD~36384, HD~52030, and HD~208742 in H$-$R diagram along with their theoretical evolution tracks (Girardi et al. 2000).
The diagram indicates the evolution of 1.5\,$M_{\odot}$ and 1.6 \,$M_{\odot}$ stars according to the stellar evolution models with Z = 0.008 (HD~52030 and HD~208742) and Z = 0.019 (HD~36384).
The stars are located at the AGB stage on the H-R diagram after having undergone the helium flash leaving the giant clump.
It is unclear what happens to the companions when they interact with the atmosphere of their expanding host stars.
%However, the companion, in the end, either will demise if engulfed by the host star or will survive if extruded from the host star.
Thus, if companions exist within 0.25 AU around HD~52030, and HD~208742 in the beginning, there are two possible scenarios for their fate. First, low-mass planets may be engulfed. Second, companions more massive than 2~$M_{\rm{Jup}}$ may be pushed out to close distance. The companions at 1.2$-$1.5 AU detected in this paper are probably impervious to the effect of the evolutionary process.
There may be unusual cases in which a short-period substellar companion survives the engulfment as its parent star evolve into a white dwarf (Maxted et al. 2006).

The current knowledge of the evolutionary tracks for evolved stars is not sufficient enough to predict the fate of these systems. We need to understand the dynamics of planetary systems around the largest evolving giants.
Furthermore, the number of large stars with planetary companions is limited. Future surveys will provide us more candidates, allowing us to better trace the shape of the companion mass function and offer us some understanding of the origin and history of these systems.

%________________________________________________________________

\acknowledgments
      BCL acknowledges partial support by Korea Astronomy and Space Science Institute (KASI) grant 2016-1-832-01. Support for MGP was provided by the KASI under the R\&D program supervised by the Ministry of Science, ICT and Future Planning and by the National Research Foundation of Korea to the Center for Galaxy Evolution Research (No. 2012-0027910). DEM acknowledges his work as part of the research activity of the National Astronomical Research Institute of Thailand (NARIT), which is supported by the Ministry of Science and Technology of Thailand. APH acknowledges grant HA 3279/8-1 from the Deutsch Forschungsgemeinschaft (DFG). SHG would like to thank the financial support by the NSFC under grant No. U1531121. This research made use of the SIMBAD database, operated at the CDS, Strasbourg, France. We thank the developers of the Bohyunsan Observatory Echelle Spectrograph (BOES) and all staff of the Bohyunsan Optical Astronomy Observatory (BOAO).

\clearpage
%-------------------------------------------------------------

 \begin{figure}
   \centering
   \includegraphics[width=8cm]{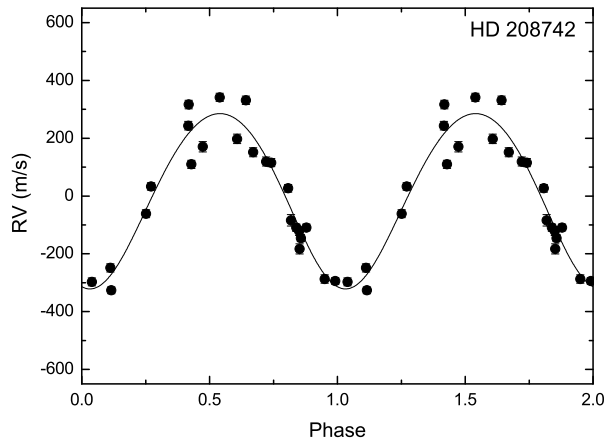}
      \caption{RV variations for HD 208742 phased with the orbital period of 602.0 days.
        }

        \label{phase3}
   \end{figure}

%
%
%-------------------------------------------------------------
\clearpage
\begin{deluxetable}{lcccc}
\tablewidth{0pt}\emph{}
\tablecaption{Stellar parameters for the stars analyzed in the present paper.
\label{tab1}}
\tablehead{
\colhead{Parameter}   & \colhead{HD 36384}  & \colhead{HD 52030}  & \colhead{HD 208742}    & \colhead{Reference} }
\startdata

    $\alpha$ (J2000)            & 05 39 43.70   & 07 05 51.67        &21 52 13.01 &  1 \\
    $\delta$ (J2000)            &+75 02 37.94   &+70 43 54.93        &+79 33 06.39&  1 \\
    Spectral type                  & M0 III        & K0 III          & K5         &  1 \\
                                   & M0 III        & M1 III          & M0.5 III   &  7 \\
    $\textit{$m_{v}$}$ (mag)       & 6.19          &  6.49           & 6.51       &  1 \\
    \emph{HIP}$_{\rm scat}$ (mag)  & 0.012         & 0.014           & 0.011      &  1 \\
    $\emph{B$-$V}$ (mag)           & 1.606         & 1.568           & 1.581      &  1 \\
    $\emph{V$-$K}$ (mag)           & 3.902         & 4.077           & 3.944      &  2 \\
    $\pi$          (mas)           &  4.64 $\pm$ 0.36   & 2.85 $\pm$ 0.54 & 2.57 $\pm$ 0.30 & 3\\
                                   &  5.10 $\pm$ 0.58   & 3.26 $\pm$ 0.46 &     --     & 4 \\
    Distance        (pc)           & 214.0 $\pm$ 16.7   &336.4 $\pm$ 63.6 & 382.8 $\pm$ 44.8& 5\\
    Age            (Gyr)           &     6.8 $\pm$ 2.7  & 5.9 $\pm$ 2.9   & 4.8 $\pm$ 2.6  & 8 \\
%   Age            (Gyr)           &     6.8 $\pm$ 2.7  & 5.9 $\pm$ 2.9   & 4.8 $\pm$ 2.6  &6\footnote{}\\
    $T_{\mathrm{eff}}$ (K)         &  3940 $\pm$ 40     & 3970 $\pm$ 30   & 3940 $\pm$ 50   & 7 \\
                                   &     3956           &  3881           & 3936            & 6 \\
    $\mathrm{[Fe/H]}$              &$-$0.16 $\pm$ 0.14  & $-$0.54  $\pm$ 0.08 &$-$0.47 $\pm$ 0.20& 7 \\
    log $\it g$                    &    1.1 $\pm$ 0.2   & 1.3 $\pm$ 0.1       &  1.1 $\pm$ 0.3   & 7 \\
    $\textit{$R_{\star}$}$ ($R_{\odot}$)        &38.4 $\pm$ 3.4  & 53.0 $\pm$ 6.1  & 57.2 $\pm$ 7.1 & 8\\
    $\textit{$M_{\star}$}$ ($M_{\odot}$)       & 1.14 $\pm$ 0.15 & 1.09 $\pm$ 0.16 &  1.18 $\pm$ 0.22 & 8\\
    $\textit{$L_{\star}$}$ [$L_{\odot}$]       & 388             &  867           & 980            & 5 \\
    $v_{\mathrm{rot}}$ sin $i$ (km s$^{-1}$)   & 4.5             &  4.1           & 5.4            & 7\\
    $P_{\mathrm{rot}}$ / sin $i$ (days)        & 440 $\pm$ 40    &  650 $\pm$ 80  & 540 $\pm$ 70   & 7\\
    $v_{\mathrm{micro}}$ (km s$^{-1}$)         & 1.5 $\pm$ 0.2 &1.5 $\pm$ 0.1   & 1.7 $\pm$ 0.2  & 7 \\

\enddata

\tablerefs{
(1) \emph{HIPPARCOS} (ESA 1997); (2) Cutri et al. (2003); (3) van Leeuwen (2007); (4)\emph{GAIA} (Gaia Collaboration et al. 2016); (5) Anderson \& Francis (2012); (6) McDonald et al. (2012); (7) This work. (8) Derived using the online tool (http://stevoapd.inaf.it/cgi-bin/param).}
%\tablenotetext{a}{See text.}

\end{deluxetable}
%

%-------------------------------------------------------------
\clearpage
\begin{deluxetable}{crr|crr}
%\tablewidth{11pc}
\tablewidth{0pt}
\tablecaption{RV measurements for HD 36384 from December 2010 to March 2016 using the BOES.
\label{tab2}}
\tablehead{\colhead{BJD} & \colhead{RV} & \colhead{$\pm \sigma$} &
            \colhead{BJD} & \colhead{RV} & \colhead{$\pm \sigma$}  \\
           \colhead{$-$2,450,000} & \colhead{m\,s$^{-1}$} & \colhead{m\,s$^{-1}$} &
           \colhead{$-$2,450,000} & \colhead{m\,s$^{-1}$} & \colhead{m\,s$^{-1}$}}
\startdata
5554.1607 &     163.6 &   12.7 &   6583.1667 &     260.7 &   12.2  \\
5843.2963 &  $-$111.9 &    9.2 &   6617.0391 &     195.1 &   10.3  \\
5934.0831 &     370.1 &   10.2 &   6739.9818 &  $-$155.4 &   11.0  \\
5963.0136 &     339.9 &    8.9 &   6922.2494 &  $-$241.7 &    9.9  \\
6015.0700 &     246.0 &   10.3 &   6965.9482 &      77.4 &    9.3  \\
6024.9427 &     328.0 &   11.6 &   7094.0446 &     185.5 &   14.7  \\
6073.9678 &      28.1 &   13.7 &   7301.1275 &  $-$209.7 &   10.7  \\
6258.1864 &  $-$275.5 &   10.0 &   7330.3056 &  $-$235.2 &    9.1  \\
6261.2126 &  $-$363.7 &    9.0 &   7378.1202 &  $-$226.3 &   11.3  \\
6287.1500 &  $-$357.9 &   17.7 &   7401.9431 &  $-$122.8 &   11.7  \\
6346.0737 &  $-$166.4 &    9.2 &   7407.9257 &  $-$104.2 &   10.3  \\
6378.1021 &   $-$38.7 &   11.0 &   7412.9413 &  $-$191.1 &   10.8  \\
6551.3145 &     253.7 &   11.4 &   7469.0464 &      50.1 &    8.9  \\
6582.2621 &     303.4 &   11.0 &             &           &         \\

\enddata
%\tablenotetext{a}{Includes scale factors described in the text.}
\end{deluxetable}
%

%-------------------------------------------------------------
\clearpage
\begin{deluxetable}{crr|crr}
%\tablewidth{11pc}
\tablewidth{0pt}
\tablecaption{RV measurements for HD 52030 from December 2010 to March 2016 using the BOES.
\label{tab3}}
\tablehead{\colhead{BJD} & \colhead{RV} & \colhead{$\pm \sigma$} &
            \colhead{BJD} & \colhead{RV} & \colhead{$\pm \sigma$}  \\
           \colhead{$-$2,450,000} & \colhead{m\,s$^{-1}$} & \colhead{m\,s$^{-1}$} &
           \colhead{$-$2,450,000} & \colhead{m\,s$^{-1}$} & \colhead{m\,s$^{-1}$}}
\startdata
5554.2573 &     46.6 &    13.2 &   6808.0071 &      78.4   &   12.4  \\
5699.0221 &    500.0 &    14.0 &   6922.2780 &  $-$432.6   &   11.6  \\
5894.2941 &     58.4 &    11.8 &   6966.2428 &  $-$111.7   &   11.5  \\
5956.1259 & $-$264.3 &    10.0 &   7025.1665 &      56.6   &   12.7  \\
6025.0571 &     60.3 &     8.9 &   7094.0781 &     291.2   &   17.1  \\
6075.0172 &    377.3 &    10.7 &   7108.1411 &     236.9   &   15.9  \\
6258.2119 &    233.0 &    10.9 &   7301.1720 &  $-$181.5   &   12.2  \\
6261.2483 &    261.2 &    13.8 &   7330.3323 &  $-$335.2   &   10.0  \\
6287.2260 &     60.7 &    14.3 &   7378.1809 &  $-$349.3   &   10.1  \\
6346.1377 & $-$121.9 &    10.2 &   7401.9600 &  $-$457.6   &   11.9  \\
6552.2975 &    129.2 &     9.1 &   7407.9424 &  $-$274.0   &   11.7  \\
6583.2662 &    347.7 &    10.5 &   7423.9772 &  $-$414.6   &    9.4  \\
6617.1340 &    287.6 &    10.0 &   7469.0631 &  $-$373.4   &   10.6  \\
6740.0213 &    291.5 &    11.0 &             &             &         \\

\enddata
%\tablenotetext{a}{Includes scale factors described in the text.}
\end{deluxetable}
%

%-------------------------------------------------------------
\clearpage
\begin{deluxetable}{crr|crr}
%\tablewidth{11pc}
\tablewidth{0pt}
\tablecaption{RV measurements for HD 208742 from February 2010 to March 2016 using the BOES.
\label{tab4}}
\tablehead{\colhead{BJD} & \colhead{RV} & \colhead{$\pm \sigma$} &
            \colhead{BJD} & \colhead{RV} & \colhead{$\pm \sigma$}  \\
           \colhead{$-$2,450,000} & \colhead{m\,s$^{-1}$} & \colhead{m\,s$^{-1}$} &
           \colhead{$-$2,450,000} & \colhead{m\,s$^{-1}$} & \colhead{m\,s$^{-1}$}}
\startdata
5251.3906 &   $-$7.6 &   12.9 &  6920.1336 &  $-$286.9 &   13.7 \\
5454.2001 &    162.9 &   16.3 &  6963.9776 &  $-$236.7 &   14.0 \\
5581.0389 & $-$114.6 &   19.9 &  6966.0079 &  $-$314.3 &   12.0 \\
5842.1071 &  $-$84.1 &   13.6 &  7147.3047 &     260.8 &   14.8 \\
6177.2612 &     14.4 &   13.3 &  7148.1940 &     334.0 &   14.8 \\
6288.0398 & $-$303.0 &   11.5 &  7300.2584 &     173.7 &   15.6 \\
6552.0723 &    108.7 &   13.8 &  7331.0875 &     142.5 &   15.6 \\
6579.0516 &    170.5 &   17.9 &  7402.0620 &   $-$84.1 &   12.6 \\
6618.9582 &    342.7 &   13.8 &  7408.9130 &   $-$96.2 &   14.5 \\
6680.3639 &    334.6 &   14.6 &  7412.9159 &  $-$119.7 &   12.3 \\
6740.3152 &    120.5 &   14.2 &  7425.9063 &   $-$82.8 &   11.8 \\
6807.1006 & $-$175.8 &   17.4 &  7469.1588 &  $-$259.9 &   14.5 \\

\enddata
%\tablenotetext{a}{Includes scale factors described in the text.}
\end{deluxetable}
%

%-------------------------------------------------------------
\clearpage
\begin{deluxetable}{lcc}
\tablewidth{0pt}
\tablecaption{Orbital parameters from Keplerian fit.
\label{tab5}}
\tablehead{
\colhead{Parameter}       & \colhead{HD 52030 b }   & \colhead{HD 208742 b}  }
\startdata
    P (days)                           & 484.3 $\pm$ 4.2     & 602.8  $\pm$ 4.9    \\
    $\it T$$_{\rm{periastron}}$ (BJD)   &2466185.7 $\pm$ 90.0 &2455690.7 $\pm$ 85.8 \\
    $\it{K}$ (m s$^{-1}$)              & 329.0  $\pm$ 21.8   & 303.0 $\pm$ 16.6  \\
    $\it{e}$                           & 0.14 $\pm$ 0.14     & 0.06 $\pm$ 0.13   \\
    $\omega$ (deg)                     & 240.3 $\pm$ 26.4    &167.1 $\pm$ 50.7   \\
    \emph{m}~sin~$\it i$ ($M_{Jup}$)   & 13.3 $\pm$ 2.3      & 14.0 $\pm$ 2.0    \\
    $\it{a}$ (AU)                      & 1.2  $\pm$ 0.1      & 1.5  $\pm$ 0.2    \\
    Slope (m s$^{-1}$ day$^{-1}$)      & ($-$1.6 $\pm$ 0.3) $\times$ 10$^{-1}$   & -- \\
    $N_{obs}$                          &  27                 &  24              \\
    rms (m s$^{-1}$)                   &  82.1               &  55.4            \\
\enddata
%\tablenotetext{a}{Star LP 608--62 is also known as BD+1\arcdeg 2341p.  We will make this footnote extra long so that it extends over two lines.}
%\tablerefs{
%(1) Barbuy, Spite, \& Spite 1985; (2) Bond 1980; (3) Carbon et al. 1987}
\end{deluxetable}

\end{document}